\documentclass[12pt]{article}
\usepackage[dvips]{color}
\usepackage{epsfig}
\usepackage{amssymb}

 1
\makeatletter
\@addtoreset{equation}{on}
\renewcommand{\theequation}{\thesection.\arabic{equation}}

\makeatother
\begin{document}
\begin{flushright}
HIP-2007-70/TH\\
KEK-TH-1216\\
December, 2007
\end{flushright}
\vspace{0.5cm}
\begin{center}
{\Large \bf
Dynamical Supersymmetry Breaking \\
\vspace{0.2cm}
from Meta-stable Vacua \\
\vspace{0.3cm}
in
an ${\cal N}=1$ Supersymmetric Gauge Theory
}

\end{center}
\vspace{8mm}
\begin{center}
\normalsize
{\large \bf  Masato Arai $^a$\footnote{masato.arai@helsinki.fi},
Claus Montonen $^a$\footnote{claus.montonen@helsinki.fi},
Nobuchika Okada $^{b}$\footnote{nobuchika.okada@kek.jp}
\vspace{2mm}\\
and
\vspace{2mm}\\
Shin Sasaki $^a$\footnote{shin.sasaki@helsinki.fi}
}
\end{center}
\vskip 1.2em
\begin{center}
{\it
$^a$ Department of Physical Sciences,
     University of Helsinki \\
 and Helsinki Institute of Physics,
 P.O.Box 64, FIN-00014, Finland \\
\vskip 1.0em
$^b$Theory Division, KEK, Tsukuba 305-0801, Japan
}
\end{center}
\vskip 0.8cm
\begin{center}
{\large Abstract}
\vskip 0.4cm
\begin{minipage}[t]{16cm}
\baselineskip=19pt
\hskip4mm
We investigate supersymmetry breaking meta-stable vacua in
 $\mathcal{N} = 2$, $SU(2) \times U(1)$ gauge theory
 with $N_f = 2$ massless flavors perturbed by the addition of small
 ${\cal N}=1$ preserving mass terms in a presence of a Fayet-Iliopoulos
 term.
We derive the low energy effective theory by using the exact results of
 ${\cal N}=2$ supersymmetric QCD and examine the effective potential.
At the classical level, the theory has supersymmetric vacua on Coulomb and
 Higgs branches.
We find that supersymmetry on the Coulomb branch is dynamically broken
 as a consequence of the strong dynamics of $SU(2)$ gauge symmetry
 while the supersymmetric vacuum on the Higgs branch
 remains.
We also estimate the lifetimes of the local minima on the Coulomb
 branch.
We find that they are sufficiently long and therefore the local vacua we
 find are meta-stable.

\end{minipage}
\end{center}
\newpage
%
%
\def\barr{\begin{eqnarray}}
\def\earr{\end{eqnarray}}

\tableofcontents

\section{Introduction}
Supersymmetry (SUSY) is the most promising and best motivated framework
 for extending the Standard Model.
However, nature turns out to be not supersymmetric
 at the electroweak scale and therefore SUSY must be broken.
The origin of the SUSY breaking is still a prime open question.
It is reasonable that SUSY is broken dynamically.
Indeed, dynamical SUSY breaking provides a natural explanation
 for the gauge hierarchy problem \cite{Witten:1981kv}.
The important fact in dynamical SUSY breaking
 is that if SUSY is not broken at tree level, it remains unbroken
 to all orders of perturbative corrections because of
 the non-renormalization theorem \cite{non}.
This implies that SUSY is dynamically broken only by non-perturbative
 effects such as instanton corrections.
Thus, understanding of gauge dynamics is crucial to study dynamical SUSY breaking.

There has been much progress in understanding
 the gauge dynamics of strongly coupled ${\cal N}=1$ SUSY field
 theory with $N_c$ color and $N_f\le N_c+1$ flavors \cite{seiberg1,seiberg2}.
The exact low energy effective superpotential can be derived
 by using the holomorphy properties of the superpotential and the gauge
 kinetic function.
This progress has triggered the discovery of many new SUSY
 breaking theories, as well
 as new techniques for establishing SUSY breaking.
One of the interesting models with dynamical SUSY breaking is the
 Izawa-Yanagida-Intriligator-Thomas model \cite{IzYa,InTh}.
In this model, an O'Raifeartaigh type sector is dynamically generated
 in the low energy superpotential.
Therefore, SUSY is spontaneously broken.
However,
 this SUSY breaking vacuum is degenerate
 {\it i.e.} there exists a pseudo flat direction.
In order to remove this degeneracy, we have to take account of
 quantum corrections for the K\"ahler potential.
In general, this is a very difficult task since the K\"ahler potential is
 not holomorphic and thus quantum corrections can be estimated at best
 by perturbative means.
Such an estimation is possible only in the ultraviolet (weak coupling)
 region of the moduli space parameterizing the pseudo flat direction
 which is far from the origin.
Therefore, the potential behavior in the infrared region remains unclear.

This situation is changed for ${\cal N}=1$ SUSY QCD with
 $N_c$ colors and $N_c + 1 \le N_f <\frac{3}{2} N_c$ flavors \cite{InSeSh}.
In this flavor region, an
 O'Raifeartaigh type model arises as the low energy effective theory of
 the magnetic dual and the effective theory is infrared free.
This is contrary to the
 Izawa-Yanagida-Intriligator-Thomas model
 where the gauge coupling strength becomes strong at low energies.
This property makes it possible to calculate perturbative
 corrections to the K\"ahler potential in the infrared region.
Indeed, in \cite{InSeSh}
 it is found that one-loop corrections to the K\"ahler potential remove
 the degeneracy of the pseudo flat direction and that there
 is a stabilized SUSY breaking vacuum at the origin of the moduli space.
In addition to this vacuum,
 there are also $N_c$ dynamically generated SUSY vacua at points far from the
 origin which are expected to exist by the argument of the Witten index.
Thus the SUSY breaking vacuum at the origin is a local vacuum.
Furthermore, the local vacuum can be long-lived compared to the age of
 the universe by choosing appropriate values of parameters in the theory.
Therefore this local vacuum is meta-stable.
Inspired by this work, further detailed researches and
 phenomenological applications have been performed
 \cite{DiMa,CsShTe,AhSe,KiOoOo,MuNo,Shih,ISS2,Fe,Al,KaShVo,AbDuJaKh,HaMa,DiMa2,AmGiMa,AmGiMa2,AmFoGiMa}.
As was mentioned above, in ${\cal N}=1$ SUSY models, one can estimate quantum
 corrections to the K\"ahler potential only in a weak coupling region
 by perturbative means.
However, in an ${\cal N}=2$ SUSY gauge theory
 one can derive the exact low energy effective action
 as was demonstrated by Seiberg and Witten \cite{s-w1,s-w2},
 using the properties of holomorphy and duality.
In \cite{ArMoOkSa},
 we studied meta-stable vacua in an $\mathcal{N} = 2$ $SU(2) \times U(1)$
 SUSY gauge theory with $N_f = 2$ massless flavors including
 a Fayet-Iliopoulos (FI) $D$-term, by using the original analysis in \cite{arai}.
Due to the FI term, the theory exhibits tree-level SUSY breaking
 on the Coulomb branch in almost all of the moduli space
 except near the origin.
Around the origin along the Coulomb branch, there is an unstable direction to
 the Higgs branch where a SUSY vacuum exists.
In this model, we demonstrated that there is a long-lived local minimum
 on the Coulomb branch in which the SUSY and $U(1)_R$ symmetry are
 dynamically broken in the non-perturbative region.
We showed that the decay rates from the local minimum to the runaway SUSY
 vacuum and also to the SUSY vacua on the Higgs branch are actually very small.
Moreover, we pointed out that massive hypermultiplets in the model
 can play the role of messenger fields in the gauge mediation scenario
 if a part of the flavor symmetry among the hypermultiplets is gauged
 and identified with the Standard Model gauge group.

It is also possible to derive the exact low energy effective action
 in the ${\cal N}=1$ theory based on the ${\cal N}=2$ theory perturbed by
 terms preserving ${\cal N}=1$ SUSY.
Assuming that the perturbation does not affect the gauge dynamics in the
 original ${\cal N}=2$ theory,
 we can use the result of the Seiberg-Witten theory.
In \cite{OoOoPa,Pa,MaOoOoPa},
 it was shown that there
 can be a meta-stable SUSY breaking vacuum
 in the Seiberg-Witten theory with terms preserving ${\cal N}=1$ SUSY.
M-theory brane configurations corresponding to these perturbed
 Seiberg-Witten theories were discussed in \cite{MaOzYa, BeGoHeSeSh}.

In this paper, we investigate a model with ${\cal N}=1$
 SUSY realizing dynamical SUSY breaking
 in meta-stable vacua.
The model we consider is an $\mathcal{N} = 2$, $SU(2) \times U(1)$ gauge theory
 with $N_f=2$ massless hypermultiplets perturbed by
 ${\cal N}=1$ preserving adjoint mass terms and a linear term (the FI $F$-term).
Although,
 in this model, only ${\cal N}=1$ SUSY is preserved by the perturbation to
 the superpotential,
 the quantum theory can be analyzed by extending the Seiberg-Witten solution,
 provided that the mass parameters $\mu_i$ and linear term parameter $\lambda$ are
 very small compared to the $SU(2)$ dynamical scale $\Lambda$.
In the classical theory of our model, there are SUSY vacua on the
 Coulomb branch and the Higgs branch.
We will show that the SUSY vacua on the Coulomb branch are dynamically broken
 as a consequence of the strong dynamics of the $SU(2)$ gauge coupling while
 the SUSY vacuum on the Higgs branch remains.
We will also show that the decay rate from the local vacua to the SUSY
 vacuum can be very small with an appropriate choice of parameters.
Therefore, we will find meta-stable SUSY breaking vacua.

The organization of this paper is as follows.
In section \ref{model}, we introduce our model and analyze the
 classical vacua.
In section \ref{quantum}, the low-energy effective action
 is derived by using exact results of $\mathcal{N} = 2$ SUSY QCD.
In section \ref{numerical}, the
 numerical analysis of the effective potential is presented.
Section \ref{decay_rate} is devoted
 to the decay rate estimation of the meta-stable SUSY vacua
 found in section \ref{numerical}.
Section \ref{conclusion} is our conclusion.
In Appendix \ref{appendix_A}, the formulas necessary for the potential
 analysis are given.

\section{The model \label{model}}
Let us first consider a tree-level Lagrangian of
 an $\mathcal{N} = 2$, $SU(2) \times U(1)$ gauge theory
 with $N_f=2$ massless fundamental flavors
 $Q$ and $\tilde{Q}$
\begin{eqnarray}
\mathcal{L}^{\mathcal{N} = 2} &=&
 \frac{1}{2 \pi} \mathrm{Im}
\left[ \mathrm{Tr} \left\{ \tau_{22} \left(
\int \! d^4 \theta \ A_2^{\dagger} e^{2 V_2} A_2 e^{ - 2 V_2}
+ \frac{1}{2} \int \! d^2 \theta \ W_2^2 \right)
\right\}
\right] \nonumber \\
& & + \frac{1}{4 \pi} \mathrm{Im} \left[
\tau_{11} \left( \int \! d^4 \theta \ A_1^{\dagger} A_1
+ \frac{1}{2} \int \! d^2 \theta \ W_1^2 \right)
\right] \nonumber \\
& & +  \int \! d^4 \theta \left[
Q_r^{\dagger} e^{2 V_2 + 2 V_1} Q^r + \tilde{Q}_r e^{- 2 V_2 - 2 V_1}
\tilde{Q}^{r \dagger} \right]
+ \sqrt{2} \left[ \int d^2 \theta \ \tilde{Q}_r (A_2 + A_1) Q^r + h.c.
\right]. \nonumber \\
\end{eqnarray}
Here, $V_2, A_2$ and $V_1, A_1$ correspond to
$SU(2)$ and $U(1)$ vector multiplets
 respectively.
The chiral superfields $Q^r_I$ and $\tilde{Q}_r^I$
 are hypermultiplets that are in the fundamental and anti-fundamental
 representations of the $SU(2)$ gauge group ($r=1,2$ is the flavor index,
 and $I=1,2$ is the $SU(2)$ color index).
The superfield strength is defined by $W_{i \alpha}
= - \frac{1}{4} \overline{D}^2 (e^{- 2 V_i} D_{\alpha} e^{2 V_i}) \
(i = 1,2)$.
The complex gauge couplings are defined by
\begin{eqnarray}
\tau_{22} = i \frac{4 \pi}{g^2} + \frac{\theta}{2 \pi},~~~~~
\tau_{11} = i \frac{4 \pi}{e^2}\,,
\end{eqnarray}
where $\tau_{22}$ corresponds to an $SU(2)$ complex gauge coupling and
 $\tau_{11}$ is a $U(1)$ gauge coupling.
The common $U(1)$ charge
 for the hypermultiplet is normalized to be 1.
The $SU(2)$ generators $T^a$ are normalized as
 $\mathrm{Tr} (T^a T^b) = \frac{1}{2} \delta^{ab}$.
The global symmetry in this theory is
 $SU(2)_{\mathrm{left}}\times SU(2)_{\mathrm{right}} \times SU(2)_R \times U(1)_R$.

Let us introduce mass and linear terms for
 the chiral superfields $A_1, A_2$,
\begin{eqnarray}
\mathcal{L}_{\mathrm{soft}} = \int \! d^2 \theta \left(
\mu_2 \mathrm{Tr} (A_2^2) + \frac{1}{2} \mu_1 A_1^2 + \lambda A_1
\right) + h.c. \label{soft}
\end{eqnarray}
These terms break $\mathcal{N} = 2$ SUSY down to $\mathcal{N} = 1$.
The dimensionful parameters $\mu_i$ can be taken
 to be real and positive without loss of generality,
 while we fix the dimensionful parameter $\lambda$
 to be real and positive, $\lambda > 0$, for simplicity.
The linear term in $A_1$ is the FI term.
In general, the FI term also appears in the $D$-term,
 but the $SU(2)_R$ symmetry allows us to take a frame so that
 it appears only in the $F$-term.
Therefore, the $SU(2)_R$ symmetry is explicitly broken down to
 its subgroup $U^\prime(1)_R$.
The superpotential (\ref{soft}) also breaks $U(1)_R$ symmetry.
The global symmetry of the theory turns out to be
 $SU(2)_{\mathrm{left}}\times SU(2)_{\mathrm{right}}\times U^\prime(1)_R$.
The scalar potential is easily derived from the Lagrangian
 $\mathcal{L} = \mathcal{L}^{\mathcal{N} = 2} + \mathcal{L}_{\mathrm{soft}}$
\begin{eqnarray}
V (a_1, a_2, q, \tilde{q}) &=& g^2 \mathrm{Tr}
 [A_2, A_2^{\dagger}]^2 + \frac{g^2}{2}
 \left( q^{\dagger}_r T^a q^r - \tilde{q}_r T^a \tilde{q}^{\dagger r} \right)^2
\nonumber \\
& & + q^{\dagger}_r [A_2, A_2^{\dagger}]q^r - \tilde{q}_r [A_2, A_2^{\dagger}]
\tilde{q}^{\dagger r} + 2 g^2 | \tilde{q}_r T^a q^r |^2
+ \frac{e^2}{2} ( q_r^{\dagger} q^r - \tilde{q}_r \tilde{q}^{\dagger r})^2
\nonumber \\
& & + 2 \left( q^{\dagger}_r |A_2 + A_1|^2 q^r + \tilde{q}_r |A_2 + A_1|^2
 \tilde{q}^{\dagger r} \right) + \sqrt{2} \mu_2 g^2 (\tilde{q}_r T^a q^r
A_2^a + h.c.) \nonumber \\
& & + \mu_2^2 g^2 A_2^{a \dagger} A_2^{a}
+ e^2 |\lambda + \mu_1 A_1 + \sqrt{2} q_r \tilde{q}^r |^2
\label{tree_potential}
\end{eqnarray}
where $A_1, A_2, q^r$ and $\tilde{q}_r$ are scalar components in the
 corresponding chiral superfields.
Without the mass and linear terms, there is a SUSY vacuum on the Coulomb branch,
\begin{eqnarray}
A_2 = \left(
\begin{array}{cc}
a_2 & 0 \\
0 & - a_2
\end{array}
\right), \quad A_1 = a_1 , \label{moduli}
\end{eqnarray}
where $a_1$ and $a_2$ are the moduli of the vacuum.
In this vacuum, the gauge symmetry is broken to $U(1)_c \times U(1)$.
When turning on the mass and the FI terms, only the following point in
 the moduli space is left as a SUSY vacuum
\begin{eqnarray}
q_r = \tilde{q}_r = 0, \quad A_2 = 0, \quad A_1 = -
 \frac{\lambda}{\mu_1}\,, \label{Coulomb-branch}
\end{eqnarray}
where the $SU(2)$ gauge symmetry is recovered.
In addition to this
 SUSY vacuum on the Coulomb branch,
 there is another SUSY vacuum on the Higgs branch given by
\begin{eqnarray}
& q_I {}^1 = \tilde{q}_I {}^{1T} =
\left(
\begin{array}{c}
u \\
v
\end{array}
\right), \quad
q_I {}^2 = \tilde{q}_I {}^{2T} =
\left(
\begin{array}{c}
v \\
- u
\end{array}
\right)\,,~~~u,v \in {\bf C}\,, &\nonumber \\
& \displaystyle
 u^2 + v^2 = \frac{-\lambda}{2 \sqrt{2}}\,,\qquad
 A_2 = A_1 = 0\,.
& \label{Higgs-branch}
\end{eqnarray}

In the following, we focus on the Coulomb branch
 and proceed to investigate the low-energy effective action.

\section{Quantum theory \label{quantum}}
\subsection{Effective action and monodromy}
\setcounter{equation}{0}
The exact low energy Wilsonian effective Lagrangian can be derived by
 integrating the action to zero momentum.
In our case, the resultant Lagrangian could be described by light
 fields, the dynamical scale, the masses $\mu_i (i=1,2)$ and the
 coefficient of the FI term $\lambda$.
However, since it is in general very difficult to implement the integration, we
 assume that $\mu_i$ and $\lambda$ are much smaller than the dynamical
 scale of the $SU(2)$ gauge interaction $\Lambda$, {\it i.e.} $\mu_i\ll \Lambda$
 and $\lambda \ll \Lambda^2$.
This setup allows us to expand the exact low energy Lagrangian
${\cal L}_{\mathrm{exact}}$ with respect to the parameters $\mu_i$ and $\lambda$ as
\begin{eqnarray}
 {\cal L}_{\mathrm{exact}}={\cal L}_{\mathrm{SUSY}}+{\cal L}_{\mathrm{soft}}
 +{\cal O}(\mu_i^2,\lambda)\,. \label{exact}
\end{eqnarray}
Here the first term ${\cal L}_{\mathrm{SUSY}}$ describes an ${\cal N}=2$
 SUSY Lagrangian containing full quantum corrections.
The second term ${\cal L}_{\rm soft}$ includes the masses and the FI
 terms in the leading order.
In the following, we consider the effective action up to the leading
 order in $\mu_i$ and $\lambda$.

First we clarify the structure of the moduli space of the theory.
As we have seen in the previous section, without
 the soft term (\ref{soft}), the theory has a moduli space parameterized
 by $a_2$ and $a_1$ on the Coulomb branch.
Except at the origin of the moduli space the gauge symmetry is broken down
 to $U(1)_c\times U(1)$.
Note that
 this $U(1)$ gauge interaction is treated as a cut-off theory \cite{arai,ArMoOkSa}.
Thus, the Landau pole $\Lambda_L$ is inevitably introduced
 in our effective theory,
 and the defining region
 of the modulus parameter $a_1$ is constrained
 to lie within the region $|a_1|< \Lambda_L$.
Because of this constraint, the defining region for the modulus parameter
 $a_2$ is also constrained to be in the same region,
 since the two moduli parameters are related to each other
 through the hypermultiplets.
We take the scale $\Lambda_L$ to be much larger than
 the dynamical scale of the $SU(2)$ gauge interaction $\Lambda$ as in
 \cite{arai,ArMoOkSa}
(The explicit scale of $\Lambda_L$
 is given at the end of this section).
This condition guarantees that the $U(1)$ gauge interaction is always weak
 in the defining region of moduli space.
Note that in our framework we implicitly assume that
 the $U(1)$ gauge interaction has no effect on the
 $SU(2)$ gauge dynamics.
This assumption will be justified
 in the following discussion concerning the monodromy transformation.

First we consider the general formulas for the effective Lagrangian
 ${\cal L}_{\mathrm{SUSY}}$.
The Lagrangian ${\cal L}_{\mathrm{SUSY}}$ is given by two parts,
 vector multiplet part $\mathcal{L}_{\mathrm{VM}}$ and
 hypermultiplet part $\mathcal{L}_{\mathrm{HM}}$;
\begin{eqnarray}
\mathcal{L}_{\mathrm{SUSY}} = \mathcal{L}_{\mathrm{VM}} + \mathcal{L}_{\mathrm{HM}}\,.
\label{effective_action}
\end{eqnarray}
The ${\cal L}_{\mathrm{VM}}$ part consists of $U(1)_c$ and $U(1)$ vector multiplets.
The $U(1)_c$ vector multiplet $(A_2, V_2)$ originates from the unbroken
 part (Cartan subalgebra)
 of the classical $SU(2)$ vector multiplet whereas $(A_1, V_1)$ belongs to
 the $U(1)$ gauge multiplet which is left unbroken from the classical level.
The effective Lagrangian for these vector multiplets is
\begin{eqnarray}
\mathcal{L}_{\mathrm{VM}} &=& \frac{1}{4 \pi} \mathrm{Im} \sum_{i,j=1}^2
\left[ \int \! d^4 \theta \ \frac{\partial \mathcal{F}}{\partial A_i} A_i^{\dagger}
+ \frac{1}{2} \int \! d^2 \theta \ \tau_{ij} W_i W_j \right], \label{VM}
\end{eqnarray}
where $\mathcal{F} = \mathcal{F} (A_2, A_1, \Lambda, \Lambda_L)$ is a prepotential
 as will be discussed below.
The effective gauge coupling $\tau_{ij}$ is defined by
\begin{eqnarray}
\tau_{ij} = \frac{\partial^2 \mathcal{F}}{\partial a_i \partial a_j}
, \quad b_{ij} \equiv \frac{1}{4 \pi} \mathrm{Im} (\tau_{ij})
\quad (i,j = 1,2).
\label{effective_coupling}
\end{eqnarray}
The hypermultiplet part $\mathcal{L}_{\mathrm{HM}}$ is
\begin{eqnarray}
\mathcal{L}_{\mathrm{HM}} &=& \int \! d^4 \theta \left[
M^{\dagger}_r e^{2 n_m V_{2D} + 2 n_e V_2 + 2 n V_1} M^r
+ \tilde{M}_r e^{- 2 n_m V_{2D} - 2 n_e V_2 - 2 n V_1} \tilde{M}^{r \dagger}
\right] \nonumber \\
& & \quad + \sqrt{2} \int \! d^2 \theta \left[ \tilde{M}_r (n_m A_{2D} + n_e A_2 + n A_1) M^r + h.c.
\right], \label{HM}
\end{eqnarray}
where $M^r, \tilde{M}_r$ are chiral superfields and
 $V_{2D}, A_{2D}$ are dual variables of $V_2, A_2$.
These hypermultiplets correspond to the light BPS dyons, monopoles and
 quarks which are specified through the appropriate quantum numbers
 $(n_e, n_m)_n$.
Here $n_e$ and $n_m$ are the electric and magnetic
 charges of $U(1)_c$, respectively, whereas  $n$ is the $U(1)$ charge.
The mass of the BPS state is specified by
\begin{eqnarray}
 M_{\mathrm{BPS}}=|n_e a_2 + n_m a_{2D} + n a_1|\,, \label{BPS}
\end{eqnarray}
where $a_{2D}$ is a scalar component of the chiral superfield $A_{2D}$.
This $\mathcal{L}_{\mathrm{HM}}$ part should be added to the effective
 Lagrangian as new degrees of
 freedom if we focus on the singular points in the moduli space.

The soft term ${\cal L}_{\mathrm{soft}}$ is given by
\begin{eqnarray}
\mathcal{L}_{\mathrm{soft}} = \int \! d^2 \theta \
\left[ \mu_2 U(A_1, A_2) + \frac{1}{2} \mu_1 A_1^2 + \lambda A_1 \right]
+ h.c., \label{soft-quantum}
\end{eqnarray}
provided that the condition $\mu_i^2, \lambda \ll \Lambda^2$ is satisfied.
Here
$U(A_2,A_1)$ is a low energy effective superfield given by
\begin{eqnarray}
U(A_2, A_1) = u (a_2, a_1) + \theta^2 F_2 \left. \frac{\partial u}{\partial a_2}
 \right|_{a_1} + \theta^2 F_1 \left. \frac{\partial u}{\partial a_1} \right|_{a_2}\,,
\end{eqnarray}
where $u$ represents a modulus field whose form in a weak coupling limit
 is $u={\rm Tr}(A_2^2)$, and $F_1$ and $F_2$ are the auxiliary
fields of $A_1$ and $A_2$, respectively.

In order to obtain an exact description of the effective Lagrangian,
 we need to find the explicit form of the prepotential $\mathcal{F}$ and
 the effective coupling $b_{ij}$.
To derive these, let us consider the monodromy
 transformations around the singular points of the moduli space.
Suppose that the moduli space is parameterized by the vector multiplet
 scalars $a_2$, $a_1$ and their duals $a_{2D}$, $a_{1D}$
 which are defined as $a_{iD}=\partial \mathcal{F}/ \partial a_i$ ($i=1,2$).
These variables are transformed into their linear combinations
 by the monodromy transformation.
In our case, the monodromy transformations form a subgroup
 of $Sp(4,\mbox{\bf R})$, which leaves the effective Lagrangian
 $\mathcal{L}_{\mathrm{VM}} + \mathcal{L}_{\mathrm{HM}}$ invariant,
 and the general formula is found to be \cite{marino3}
\begin{eqnarray}
 \left(\begin{array}{c}
       a_{2D} \\  a_{2}  \\  a_{1D} \\  a_1  \end{array}  \right)
 \rightarrow
 \left(\begin{array}{c}
       \alpha a_{2D}+\beta a_2+p a_1   \\
       \gamma a_{2D}+\delta a_2+q a_1  \\
       a_{1D}+p(\gamma a_{2D}+\delta a_2)
        -q(\alpha a_{2D}+\beta a_2)-pqa_1\\ a_1
\end{array}  \right)\,,
  \label{eq:mono}
\end{eqnarray}
where
$
\left(\begin{array}{cc}
\alpha & \beta  \\ \gamma & \delta \end{array} \right)
 \in SL(2,\mbox{\bf Z})
$
 and $p,q \in\mbox{\bf Q}$.
Note that this monodromy transformation for the combination
 $(a_{2D},a_{2},a_1)$ is exactly the same as that
 for $\mathcal{N} = 2$ $SU(2)$ SUSY QCD with $N_f = 2$ massive quark hypermultiplets,
 if we regard $a_1$ as the common mass $m$ of the hypermultiplets
 such that $m= \sqrt{2} a_1$.
This fact means that the $U(1)$ gauge interaction part
 only plays the role of the mass term for the $SU(2)$ gauge dynamics.
This observation is consistent with our assumptions.
On the other hand, the $SU(2)$ dynamics plays an important role
 for the $U(1)$ gauge interaction through the hypermultiplet part,
 as can be seen from the transformation law of $a_{1D}$.
This monodromy transformation is also used to derive dual variables
 associated with the BPS states.
As a result, the prepotential of our theory turns out to be
 essentially the same as the result in Ref.~\cite{s-w2}
 with the additional relation $m=\sqrt{2}A_1$,
\begin{eqnarray}
\mathcal{F} (A_2,A_1,\Lambda, \Lambda_L)
 = \mathcal{F}_{SU(2)}^{(SW)} (A_2, m,\Lambda)
    \Bigg{|}_{m=\sqrt{2}A_1}+C A_1^2\,,
    \label{eq:pre}
\end{eqnarray}
where the first term on the right hand side
 is the prepotential of ${\cal N}=2$ SUSY QCD
 with hypermultiplets having the same mass $m$,
 and $C$ is a free parameter.
The freedom of the parameter $C$ is used to determine
 the scale of the Landau pole
 relative to the scale of the $SU(2)$ dynamics.
For instance, taking $C=4\pi i$ leads to the value of the Landau pole
 $\Lambda_L\sim 10^{17-18} \Lambda$ (for more detail, see Appendix
 \ref{appendix_A} and also \cite{arai,ArMoOkSa}).

Now that we have obtained the explicit form of the effective
 Lagrangian, let us move on to the analysis of the potential.

\subsection{Effective potential}
We can write down the effective potential from
 (\ref{exact}) with (\ref{VM}), (\ref{HM}) and (\ref{soft-quantum}).
After using the equation of motion for the auxiliary fields
 $D_i, F_i, F_{M}$ and $F_{\tilde{M}} \ (i = 1,2)$
 of the superfields $V_i, A_i, M$ and $\tilde{M}$,
 the effective potential is written as
\footnote{
 We assume that the potential is described by the proper variables
 associated with the light BPS states.
For example, the variable $a_2$ is understood implicitly as $-a_{2D}$
 when we consider the effective potential for the monopole.}
\begin{eqnarray}
V = b_{ij} F_i F_i^{\dagger} + \frac{1}{2} b_{ij} D_i D_j
+ |F_{M}|^2 + |F_{\tilde{M}}|^2. \label{effective_potential_auxiliary}
\end{eqnarray}
Here
\begin{eqnarray}
& & D_1 = \frac{b_{12} - n b_{22}}{\det b} (|M^r|^2 - |\tilde{M}_r|^2),
\label{D1} \\
& & D_2 = \frac{-(b_{11} - n b_{12})}{\det b} (|M^r|^2 - |\tilde{M}_r|^2),
\label{D2} \\
& & F_1 = \frac{-1}{\det b} \left[ \sqrt{2} M^{\dagger}_r \tilde{M}^{r \dagger}
(n b_{22} - b_{12}) + X^{\dagger} \right],
\label{F1} \\
& & F_2 = \frac{1}{\det b} \left[ \sqrt{2} M^{\dagger}_r \tilde{M}^{r \dagger}
(n b_{12} - b_{11}) + Y^{\dagger} \right],
\label{F2} \\
& & F_{M} = - \sqrt{2} (a_2^{\dagger} + n a_1^{\dagger})
\tilde{M}^{r \dagger},
\label{FM} \\
& & F_{\tilde{M}} = - \sqrt{2} (a_2^{\dagger} + n a_1^{\dagger}) M^{\dagger}_r,
\label{FtildeM}
\end{eqnarray}
where $\det b=b_{11}b_{22}-b_{12}^2$ and
\begin{eqnarray}
&
 |M^r|^2 = M^r M^{\dagger}_r, \quad |\tilde{M}_r|^2 = \tilde{M}_r
\tilde{M}^{r \dagger},&\\
& \displaystyle
 X = b_{22}(\lambda + \mu_1 a_1) + \mu_2 \left(b_{22} \frac{\partial u}{\partial a_1}
- b_{12}\frac{\partial u}{\partial a_2}\right),& \\
&  \displaystyle
 Y = b_{12}(\lambda + \mu_1 a_1) + \mu_2 \left(-b_{11} \frac{\partial u}{\partial a_2}
+ b_{12} \frac{\partial u}{\partial a_1}\right).&
\end{eqnarray}
After plugging the solution (\ref{D1})-(\ref{FtildeM})
 into (\ref{effective_potential_auxiliary}),
 the potential is rewritten in terms of $a_1, a_2$,  $M, \tilde{M}$.
The result is
\begin{eqnarray}
V(a_2, a_1, M, \tilde{M}) &=& S \left[
(|M^r|^2 - |\tilde{M}_r|^2)^2 + 4 |M^r \tilde{M}_r|^2 \right]
+ 2 T (|M^r|^2 + |\tilde{M}_r|^2 ) + U \nonumber \\
& & \qquad + \frac{\sqrt{2}}{\det b} \left[ M_r^{\dagger} \tilde{M}^{r \dagger}
(n X - Y) + h.c. \right]. \label{effective_potential_STU}
\end{eqnarray}
Here we have defined
\begin{eqnarray}
S &\equiv& \frac{1}{2 b_{22}} + \frac{(b_{12} - n b_{22})^2}{2 b_{22} \det b}, \\
T &\equiv& |a_2 + n a_1|^2, \\
U &\equiv& \frac{1}{\det b} \left[
b_{22} {\Bigg |} (\lambda + \mu_1 a_1) + \mu_2 \frac{\partial u}{\partial a_1}  \right|^2
 + b_{11} \mu_2^2 \left| \frac{\partial u}{\partial a_2} \right|^2
  \nonumber \\
& &- \left\{ (\lambda + \mu_1 a_1) \mu_2 b_{12}
 \frac{\partial u^{\dagger}}{\partial a_2^{\dagger}}
 + \mu_2^2 b_{12} \frac{\partial u^{\dagger}}{\partial a_1^{\dagger}}
 \frac{\partial u}{\partial a_2}  + h.c. \right\} {\Bigg ]}. \label{U}
\end{eqnarray}
Let us consider the stationary conditions in the hypermultiplet
directions $M$ and $\tilde{M}$,
\begin{eqnarray}
0 &=& \frac{\partial V}{\partial M^{\dagger}}
= S \left[ 2 (|M|^2 - |\tilde{M}|^2) M + 4 (M \tilde{M}) \tilde{M}^{\dagger} \right]
+ 2 T M + \frac{\sqrt{2}}{\det b} \tilde{M}^{\dagger} (n X -
Y), \label{M_stational} \\
0 &=& \frac{\partial V}{\partial \tilde{M}^{\dagger}} =
S \left[ 2 (|M|^2 - |\tilde{M}|^2) (- \tilde{M}) + 4 (M\tilde{M}) M^{\dagger}
\right]
 + 2 T \tilde{M} + \frac{\sqrt{2}}{\det b} M^{\dagger}
(n X - Y), \label{tildeM_stational}
\end{eqnarray}
where we have suppressed the indices for simplicity.
From these equations, we find
\begin{eqnarray}
 2 \left[ S (|M|^2 + |\tilde{M}|^2) + T \right] (|M|^2 - |\tilde{M}|^2)
 = 0.
\end{eqnarray}
Since $ S > 0$ and $T > 0 $, we obtain the condition $|M| = |\tilde{M}|$.
This allows us to re-express $M$ and $\tilde{M}$ as
\begin{eqnarray}
& |M| = |\tilde{M}| \equiv \mathcal{M}, &\nonumber \\
&M \equiv \mathcal{M} e^{i \vartheta},\quad
 \tilde{M} \equiv \mathcal{M} e^{i \tilde{\vartheta}}. \label{Mphase}
\end{eqnarray}
Using the condition $|M| = |\tilde{M}|$, we find
\begin{eqnarray}
0 &=& \frac{\partial V}{\partial M^{\dagger}}
= 4 S \mathcal{M}^3 e^{i \vartheta} + 2 T \mathcal{M} e^{i \vartheta}
+ \frac{\sqrt{2}}{\det b} \mathcal{M} e^{- i \tilde{\vartheta}} (n X - Y), \label{stational1} \\
0 &=& \frac{\partial V}{\partial M}
= 4 S \mathcal{M}^3 e^{- i \tilde{\vartheta} }
 + 2 T \mathcal{M} e^{ - i \tilde{\vartheta} }
 + \frac{\sqrt{2}}{\det b} \mathcal{M} e^{i \vartheta}
(n X - Y)^{\dagger}\,. \label{stational2}
\end{eqnarray}
These equations fix the phases ${\vartheta}$ and $\tilde{\vartheta}$ as
\begin{eqnarray}
e^{i (\vartheta + \tilde{\vartheta})} =
 \pm \left[ \frac{n X - Y}{(n X - Y)^{\dagger}} \right]^{\frac{1}{2}}. \label{phase}
\end{eqnarray}
Substituting this solution into (\ref{stational2}) gives
\begin{eqnarray}
\mathcal{M} \left[ 4 S \mathcal{M}^2 + 2 T \pm \frac{\sqrt{2}}{\det b} |n X - Y| \right] = 0\,.
\end{eqnarray}
We find the following solution for the above equation
\begin{eqnarray}
& & {\rm 1)}~\mathcal{M} = 0\,, \label{solution1}\\
& & {\rm 2)}~\mathcal{M}^2 = - \frac{T}{2S} \mp \frac{\sqrt{2}}{4 S \det b} |n X
- Y|\,. \label{solution2}
\end{eqnarray}
The positivity of $\mathcal{M}^2$ requires us to take the plus sign in (\ref{solution2}).
Corresponding to the solutions (\ref{solution1}) and (\ref{solution2}),
 we have the following forms of the scalar potentials
\begin{eqnarray}
 &&{\rm 1)}~V(a_2(u,a_1),a_1)=U\,,\label{potential1}\\
 &&{\rm 2)}~V(a_2(u,a_1),a_1)=U-4S\mathcal{M}^4\,. \label{potential2}
\end{eqnarray}
The solution (\ref{potential2}) where the light hypermultiplet acquires
 a vacuum expectation value is energetically favored because $S>0$.
The potential minimum is expected to emerge at the singular points on the
 moduli space since the hypermultiplets appear in the theory as the
 light BPS states there.
In addition, the solutions are stable in the ${\cal M}$
 direction.
This is because they are unique solutions and have a lower energy than
 (\ref{solution1}).
Furthermore the potential
 is dominated by $\mathcal{M}^4$ term with a positive coefficient for
 a large value of ${\cal M}$.
On the other hand, the solution (\ref{potential1}) describes the behavior
 away from the singular points.
It smoothly connects with the solution (\ref{potential2}).

In the next section we examine the effective potential
 (\ref{potential2}) numerically.
The potential is a function of the periods $a_{2D}, a_2$ and the
 effective couplings $\tau_{ij}$.
 In order to perform the analysis of the potential, we need their explicit forms.
Their detailed derivation was given in \cite{arai,ArMoOkSa}.
In the Appendix \ref{appendix_A}, we assemble these forms and also
 display other necessary formulas for the analysis of the potential.

\section{Numerical analysis \label{numerical}}
\subsection{Singular points}
\setcounter{equation}{0}

As explained in the previous section, the minimum is expected to appear
 at the singular point since it is energetically
 favored due to the non-zero condensation of the light
 BPS state (see eq. (\ref{potential2})).
Thus, let us first investigate the singular points
 before analyzing the effective potential at the singular point.

\begin{figure}
\begin{center}
\leavevmode
  \epsfysize=4.7cm
  \epsfbox{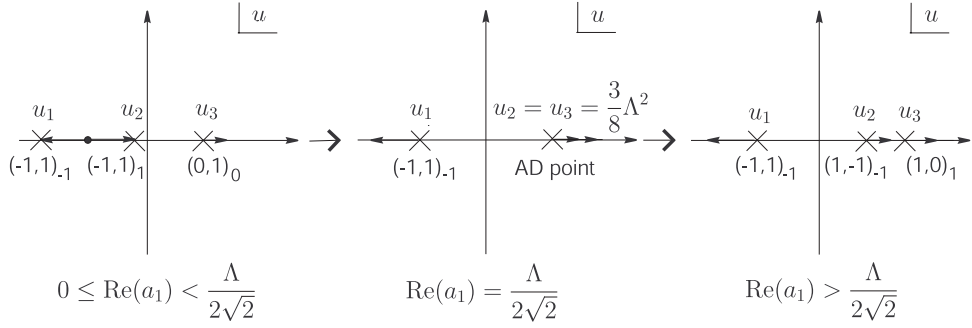} \\
\caption{Flow of the singular points as ${\rm Re}(a_1)$ increases with
 ${\rm Im}(a_1)=0$.}
\label{evo2}
\end{center}
\end{figure}

The singular points on the moduli space are determined
 by the cubic polynomial \cite{s-w2}.
The solutions of the cubic polynomial give the positions of the singular
 points in the $u$-plane.
In the case $N_c = N_f=2$ with a common hypermultiplet mass $m$,
 which is regarded as the modulus $\sqrt{2} a_1$ here,
 the solution is easily obtained as
\begin{eqnarray}
u_1=-m\Lambda-\frac{\Lambda^2}{8}{\Bigg |}_{m=\sqrt{2}a_1}\,, \;
u_2=m\Lambda-\frac{\Lambda^2}{8}{\Bigg |}_{m=\sqrt{2}a_1}\,, \;
u_3=m^2+\frac{\Lambda^2}{8}{\Bigg |}_{m=\sqrt{2}a_1}\,. \label{descriminant}
\end{eqnarray}
Let us first consider the case $\mathrm{Im} (a_1) = 0$.
The flow of the singular points with respect to
 $a_1$ is sketched in Fig.~\ref{evo2}.
For $a_1=0$, the singular points appear at $u_1=u_2=-\Lambda^2/8$
 and $u_3=\Lambda^2/8$, which correspond to the dyon and the monopole
 BPS states with quantum numbers $(n_e,n_m)_n=(-1,1)_0$ and $(0,1)_0$,
 respectively.
When switching on $a_1$, the degenerate
 dyon point splits into two singular points $u_1$ and $u_2$,
 whose BPS states are dyons with quantum numbers
 $(-1,1)_{-1}$(left dyon) and $(-1,1)_1$ (right dyon), respectively.
As $a_1$ is increasing, these singular points, $u_1$ and $u_2$,
 are moving to the left and the right on the real $u$-axis.
The two singular points, $u_2$ and $u_3$, collide and coincide
 at the so-called Argyres-Douglas (AD) point \cite{ArDo}
 ($u=\frac{3\Lambda^2}{8}$) for
 $a_1=\Lambda/(2\sqrt{2})$,
 where it is believed that the theory becomes superconformal.
As $a_1$ increases further, there appear two singular points
 $u_2$ and $u_3$ again, and the quantum numbers of the corresponding BPS states,
 $(-1,1)_1$ at $u_2$ and $(0,1)_0$ at $u_3$, change into
 $(1,1)_{-1}$(right dyon) and $(1,0)_1$ (quark), respectively.
The singular point $u_3$
 is then moving away to the right faster than $u_2$.
Note that for ${\rm Im}(a_1) =0$,
 it is not necessary to consider the case for $a_1<0$,
 since the result for $a_1 < 0$ can be obtained
 by exchanging $u_1 \leftrightarrow u_2$,
 as can be seen from the first two equations in eq.~(\ref{descriminant}).

\subsection{Numerical calculation}
Let us examine the effective potential (\ref{potential2})
 numerically.
Since the potential minimum appears at the singular point, it is
 sufficient to investigate the behavior of the effective potential around the
 singular point.
This consideration simplifies the numerical calculations.
The singular point is specified by (\ref{descriminant}) and thus the
 potential at the singular point becomes just a function of $a_1$.
In the following we investigate the effective potential at some fixed
 values of $a_1$ and see how the minimum appears at the singular point.
Then we examine the evolution of the minimum by varying $a_1$.
In the whole numerical analysis, we take $\Lambda=2\sqrt{2}$.
The values of $\mu_i$ and $\lambda$ will be taken so that the conditions
 $\mu_i\ll \Lambda$ and $\lambda \ll \Lambda^2$ are satisfied.

\begin{figure}[htb]
\begin{center}
\includegraphics[scale=.80]{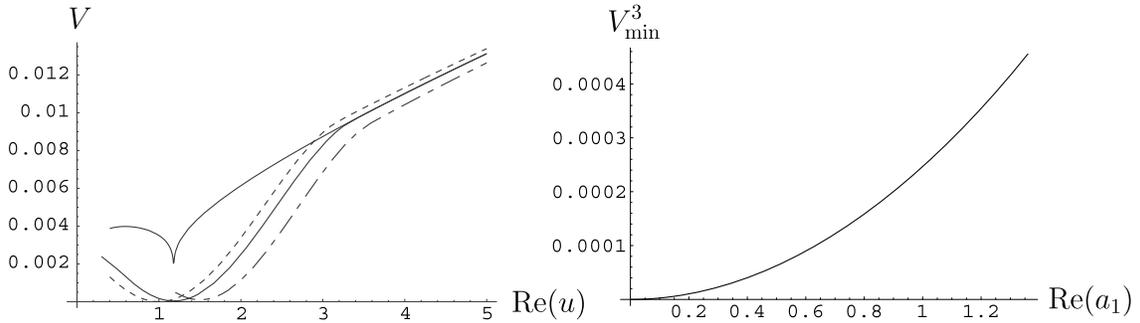}
\end{center}
\caption{The left figure shows the plots
 of the potential around the monopole singular point
 as a function of real $u$ with
 $a_1=0.2$ (dotted), $0.3$ (solid) and $0.4$ (dash-dotted).
For the case $a_1=0.3$, the potential is shown both with (bottom curve) and without (upper curve)
 condensation.
The right figure shows the evolution of the potential minimum at the
 monopole singular point $V_{\mathrm{min}}^3$ as a function of real $a_1$.}
\label{mono-evo}
\end{figure}

Since the singular points in the moduli space exhibit
 different behaviors according to the value of $a_1$, let us
 separate the $a_1$ region into three parts, namely,
 (i) $0 \le \mathrm{Re} (a_1) < \frac{\Lambda}{2 \sqrt{2}}$,
 (ii) $\mathrm{Re} (a_1) = \frac{\Lambda}{2 \sqrt{2}}$,
 (iii) $\mathrm{Re} (a_1) > \frac{\Lambda}{2 \sqrt{2}}$.
In each region, we also consider the $\mathrm{Im}(a_1)$ direction.
Let us first analyze the case $\mu_1=\lambda = 0$.
In this case, the soft term ${\cal L}_{\mathrm{soft}}$ is simply
\begin{eqnarray}
 {\cal L}_{\mathrm{soft}}=\mu_2^2\int d^2\theta U(A_2,A_1)+h.c.\,. \label{soft-mu2}
\end{eqnarray}
Note that now there exists symmetry between two BPS states
 at the singular points $u_1$ and $u_2$ for the region (i).
They are invariant
 under the interchanges $a_1\leftrightarrow -a_1$ and $n\leftrightarrow
 -n$ (see (\ref{BPS}) and (\ref{descriminant})).
\\
\\
\underline{(i) $0 \le \mathrm{Re} (a_1) < \frac{\Lambda}{2 \sqrt{2}}$}
\\
\\
In this region, there are two dyons corresponding to $u_1$ (left dyon)
 and $u_2$ (right dyon)
 and a monopole corresponding to $u_3$.
As anticipated in the discussion in section
 \ref{quantum}, there are three potential minima
 at these singular points.
The left figure in Fig. \ref{mono-evo} shows the effective potential
 around the monopole singular point along the real $u$-axis
 for several fixed values of $a_1$ with $\mu_2=0.1$.
There potential has a minimum at the singular point.
%
%
The upper solid curve shows the potential without
 the monopole condensation (\ref{potential1}) and the bottom
 solid curve includes the condensation (\ref{potential2}) for $a_1=0.3$.
The cusps in the potential are smoothed out by introducing BPS states.
It shows that the BPS state enjoys correct degrees of freedom.
The other curves are plots for $a_1=0.2$ (dotted) and $a_1=0.4$ (dash-dotted).
Note that the energy of the potential minimum is not zero except $a_1=0$
 as we will show below.
Now we examine how this minimum evolves as $a_1$ varies.
The right figure in Fig. \ref{mono-evo} shows the evolution of the
 potential minimum at the monopole singular points $V_{\mathrm{min}}^3$
 as a function of $a_1$ with $\mu_2=0.1$ and ${\rm Im}(a_1)=0$.
As $a_1$ is decreasing, $V_{\mathrm{min}}^3$ monotonically
 decreases and $V_{\mathrm{min}}^3=0$ at $a_1=0$.
%
%
\begin{figure}[htb]
\begin{center}
\includegraphics[scale=.8]{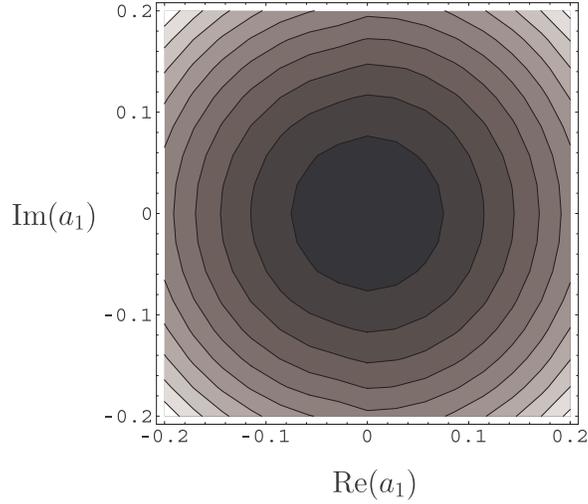}
\end{center}
\caption{The contour plot at monopole singular point as a function of
 complex $a_1$.}
\label{contour}
\end{figure}
The behavior of $V_{\mathrm{min}}^3$ for complex values of $a_1$
 is shown as the contour plot in Fig. \ref{contour}.
The dark (light) color shows lower (higher) value of the effective
 potential.
Thus, the potential minimum $V_{\mathrm{min}}^3$ is a SUSY vacuum at
 ${\rm Re}(a_1)={\rm Im}(a_1)=0$.

\begin{figure}
\begin{center}
\includegraphics[scale=.8]{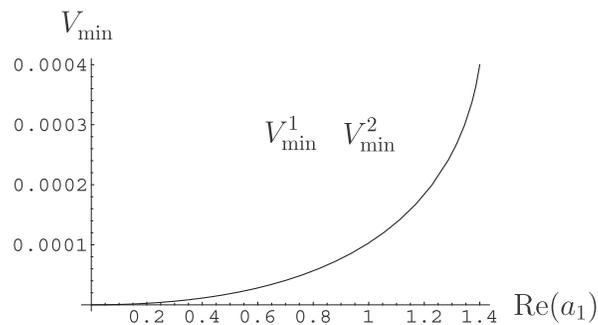}
\end{center}
\caption{Plots for the evolution of the minima at the left and the
 right dyon singular points.}
\label{dyon-evo}
\end{figure}

A similar analysis can be performed for the other singular points.
The evolution of the potential energy
 at the right dyon singular
 point $V_{\rm min}^2$ as a function of ${\rm Re}(a_1)$ with $\mu_2=0.1$
 and ${\rm Im}(a_1)=0$ is shown in Fig. \ref{dyon-evo}.
The evolution of the effective potential at the left
 dyon singular point $V_{\rm mim}^1$
 has the same behavior as $V_{\rm min}^2$
 since the singular points at $u_1$ and $u_2$ get interchanged under
 $a_1\leftrightarrow -a_1$ and $n\leftrightarrow -n$ as mentioned in the
 previous subsection.
%
%

We have seen that the theory has two SUSY vacua at the monopole and
 (degenerate) dyon singular points at $a_1=0$.
This result can be understood from the fact that the moduli structure of
 ${\cal L}_{\mathrm{SUSY}}$ for vanishing $a_1$ is the same as the one
 of ${\cal N}=2$ $SU(2)$ theory with $N_f=2$ massless flavors.
Recall that ${\cal L}_{\mathrm{SUSY}}$ includes
 the prepotential ${\cal F}_{SU(2)}^{(SW)}$ in (\ref{eq:pre}) which
 describes the moduli space $u$.
For vanishing mass ($a_1=0$), this prepotential is the same as that of
 ${\cal N}=2$ $SU(2)$ with $N_f=2$ massless flavors
 which has a ${\bf Z}_2$ symmetry, $u\leftrightarrow -u$ \cite{s-w2}.
The soft SUSY breaking term ${\cal L}_{\mathrm{soft}}$ with
 $\mu_1=\lambda=0$ has the effect of lifting up the potential in all of
 moduli space except at the monopole and the dyon singular points
 for $a_1=0$.
The remaining vacua exhibit the ${\bf Z}_2$ symmetry.

Below we shall show that when $\mu_1$ and $\lambda$ are switched on,
 SUSY at these dyon and monopole points is broken dynamically.
\\
\\
\underline{(ii) $ \mathrm{Re} (a_1) = \frac{\Lambda}{2 \sqrt{2}}$}
\\
\\
At the point $a_1 = \frac{\Lambda}{2 \sqrt{2}}$ (AD point),
 the two potential minima at the right dyon singular
 point $V^2_{\mathrm{min}}$ and
 at the monopole singular point $V^3_{\mathrm{min}}$ coincide.
As we have mentioned, it is expected that the theory becomes
 superconformal.
However, we have no knowledge of the correct description
 of the theory at this point.
\\
\\
\underline{(iii) $ \mathrm{Re} (a_1) > \frac{\Lambda}{2 \sqrt{2}}$}
\\
\\
%
%
\begin{figure}
\begin{center}
\includegraphics[scale=.8]{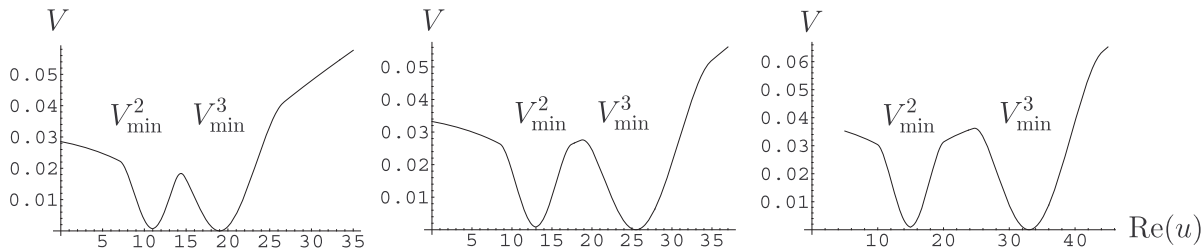}
\end{center}
\caption{Plots of the potential along real
 $u$-axis with the values $a_1=4,~4.5,~5$ from left to right. }
\label{quark-evo}
\end{figure}
%
%
\begin{figure}
\begin{center}
\includegraphics[scale=.8]{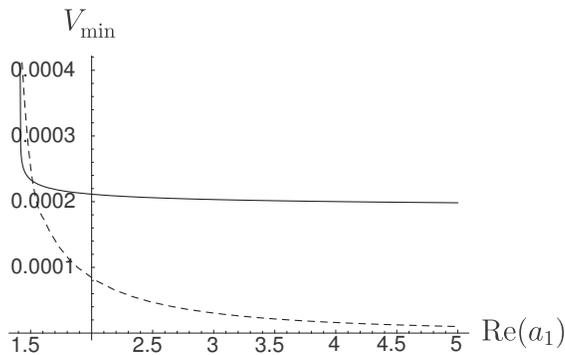}
\end{center}
\caption{The evolutions of the potential energies
 $V_{\mathrm{min}}^2$(solid)
 and $V_{\mathrm{min}}^3$(dashed).}
\label{rdyon-quark-evo}
\end{figure}
%
%
\begin{figure}[htb]
\begin{center}
\includegraphics[scale=.9]{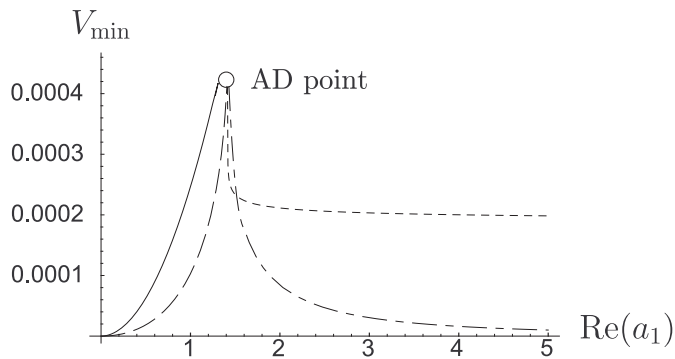}
\end{center}
\caption{Global structure of vacuum.
Solid and dashed curves show the evolutions of the potential energies at the
 monopole and left(right) dyon points for $0\le {\rm Re}(a_1)<\Lambda/(2\sqrt{2})$.
The potential energies at the right dyon(dotted)
 and quark(dash-dotted) points for ${\rm Re}(a_1)>\Lambda/(2\sqrt{2})$ are also plotted.}
\label{global_structure_potential}
\end{figure}
In this region, there are again three singular points and
 correspondingly three potential minima, $V_{\mathrm{min}}^1$ at
 $u_1$ (left dyon), $V_{\mathrm{min}}^2$ at $u_2$ (right dyon) and
 $V_{\mathrm{min}}^3$ at $u_3$ (quark).
Fig. \ref{quark-evo} shows
 the effective potential along the real $u$ axis
 around the right dyon and the quark singular
 points for several values of $a_1$.
We note that the energy at the potential minimum is not
 zero expect certain point.
The evolutions of the two minima
 at the right dyon and the quark singular points $V_{\mathrm{min}}^2$ and
 $V_{\mathrm{min}}^3$
 are depicted in Fig. \ref{rdyon-quark-evo}.
The potential energies $V_{\mathrm{min}}^2$ and
 $V_{\mathrm{min}}^3$ approach zero as $a_1$ is decreasing,
 while the evolution of the potential energy at the singular point $u_1$
 is the same as for $0\le{\rm Re}(a_1)<\Lambda/(2\sqrt{2})$.
Thus, there are runaway directions along the flow of the right dyon
 and the quark singular points.
We can find the same global structure along the flows of these two
 singular points for general complex $a_1$ values.

\begin{figure}[htb]
\begin{center}
\hspace{20mm}\includegraphics[scale=.7]{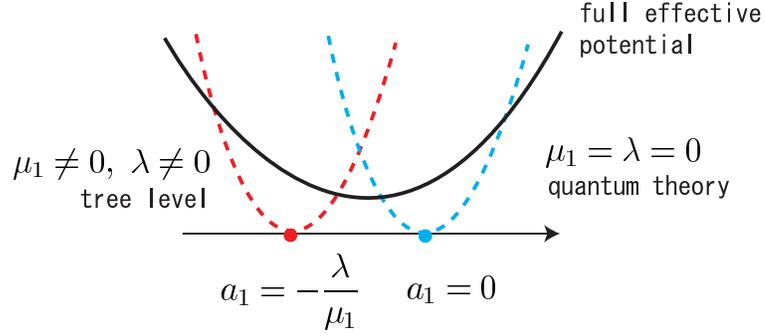}
\end{center}
\caption{Schematic picture of SUSY breaking mechanism}
\label{IYIT}
\end{figure}

The evolutions of the potential energies according the flows of the
 singular points along the real $u$-axis are simultaneously plotted in
 Fig. \ref{global_structure_potential}.
The theory has SUSY vacua at $a_1=0$ and infinity, and no
 (local) SUSY breaking vacuum.
However, this analysis gives us an important piece of information.
In the presence of the soft term (\ref{soft-mu2}),
 the gauge dynamics favors the monopole and the dyon points at $a_1=0$
 as SUSY vacua besides the runaway vacua.
It implies that if we can add certain terms to (\ref{soft-mu2})
 which produce a vacuum at a point different from $a_1=0$
 at the classical level, SUSY is dynamically broken as a consequence of
 the discrepancy of SUSY conditions between the classical and the
 quantum theories.
Actually, turning on the mass $\mu_1$ and the FI parameter
 $\lambda$ realizes such a situation.
In this case, the classical vacuum is at $a_1=-\lambda/\mu_1$,
 different from the point $a_1=0$ which the dynamics favors.
A resultant SUSY breaking vacuum is realized at
 non-zero value of $a_1$.
This is
 very similar to the SUSY breaking mechanism discussed in the
 Izawa-Yanagida-Intriligator-Thomas model in ${\cal N}=1$ SUSY gauge
 theory \cite{IzYa, InTh}.
We show a schematic picture of our situation in Fig. \ref{IYIT}.

\begin{figure}[htb]
\begin{center}
\includegraphics[scale=.9]{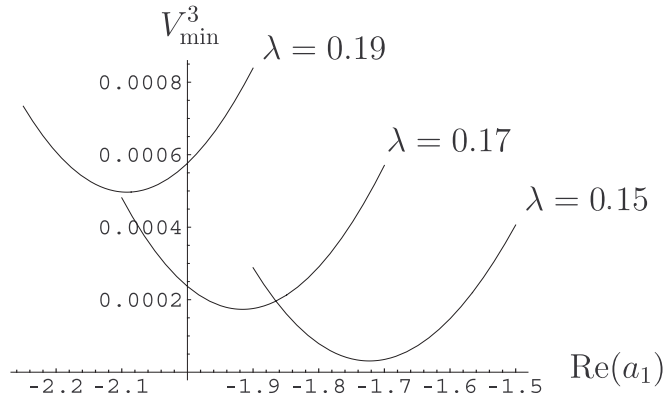}
\end{center}
\caption{Local SUSY breaking minimum at the monopole singular point for
 $\mu_1=\mu_2=0.1$ and $\lambda=0.15,0.17,0.19$ from bottom to top.}
\label{monopole_minimum}
\end{figure}
Let us see in detail how this works for non-zero values of $\mu_1, \mu_2$ and $\lambda$.
First we investigate the case $0\le {\rm Re}(a_1)<\Lambda/(2\sqrt{2})$.
Fig. \ref{monopole_minimum} shows the evolution of the potential energies
 at the monopole point $V_{\mathrm{min}}^3$ for several values of
 $\lambda$ as a function of ${\rm Re}(a_1)$ with $\mu_1=\mu_2=0.1$.
The potential minimum is no longer realized at $a_1=0$, but the location
 is shifted to negative values of ${\rm Re}(a_1)$ as is expected from
 the discussion in the previous paragraph (see also Fig. \ref{IYIT}).
Furthermore, the potential energy has a non-zero value and therefore
 SUSY is dynamically broken.
The potential energy becomes large as $\lambda$ is increasing.
This is expected from the fact that the effective potential
 behaves as $V\sim \lambda^2$ (see (\ref{U}) and (\ref{potential2})).
We also find that the potential minimum at the monopole point is
 stable for general complex values of $a_1$ (for the $\mu_1=\lambda=0$ case,
 see Fig. \ref{contour}).

The same situation occurs at the degenerate dyon singular point.
Recall that for vanishing $\mu_1$ and $\lambda$ the theory has vacua
 at the degenerate dyon point $(u,a_1)=(-\Lambda^2,0)$
 and at the monopole point $(u,a_1)=(\Lambda^2,0)$.
These two vacua are transformed into each other under the ${\bf Z}_2$ symmetry
 $u\leftrightarrow -u$.
Since turning on $\mu_1$ and $\lambda$ does not break
 this symmetry, it is also expected that SUSY is dynamically broken
 at the degenerate dyon point as it is shifted towards the negative direction of
 ${\rm Re}(a_1)$.
Therefore we now have two SUSY breaking minima at the degenerate dyon
 point and at the monopole point.

\begin{figure}[htb]
\begin{center}
\includegraphics[scale=.8]{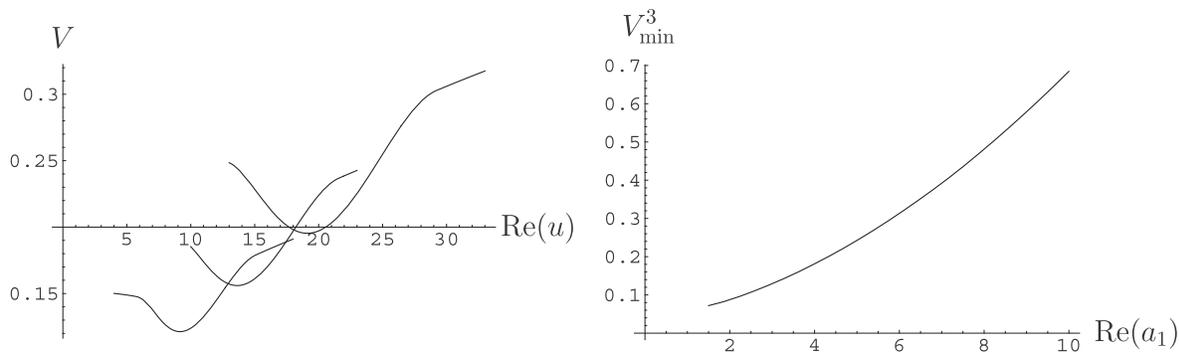}
\end{center}
\caption{The left figure shows the effective potential around the quark singular
 point with $a_1=2,~2.5,~3$ from the bottom to the top.
The right figure shows the evolution of the potential minimum
 at the quark singular point as a function of real $a_1$
 with $\mu_1=\mu_2=0.1$ and $\lambda=0.15$.}
\label{quark-lambda}
\end{figure}

In order to see the global structure of the effective potential
 we also need to investigate the potential for ${\rm Re}(a_1)> \Lambda/(2\sqrt{2})$.
Fig. \ref{quark-lambda} shows the potential energy around the quark
 singular point $V_{\mathrm{min}}^3$ as a function of ${\rm Re}(u)$
 and the evolution of $V_{\mathrm{min}}^3$ as a function of ${\rm
 Re}(a_1)$ with $\mu_1=\mu_2=0.1$ and $\lambda=0.15$.
Notably, the potential energy becomes large as $a_1$ is increasing.
This behavior is completely different from the one of the $\mu_1=\lambda=0$ case.
This difference can be understood from the classical potential
 (\ref{tree_potential}).
Since we are considering the Coulomb branch, substitute (\ref{moduli})
 with $q=\tilde{q}=0$ into (\ref{tree_potential}).
Then we obtain
\begin{eqnarray}
 V=2\mu_2^2 g^2a_2^2 + e^2|\lambda+\mu_1 a_1|^2\,. \label{cl-pot}
\end{eqnarray}
For large values of $a_1$ the term $e^2\mu_1^2 |a_1|^2$ is dominant.
Therefore the potential energy increases monotonically with growing $a_1$.
We find that the potential energies at the left and right dyon singular
 points also have the same structure.

\begin{figure}[htb]
\begin{center}
\includegraphics[scale=.6]{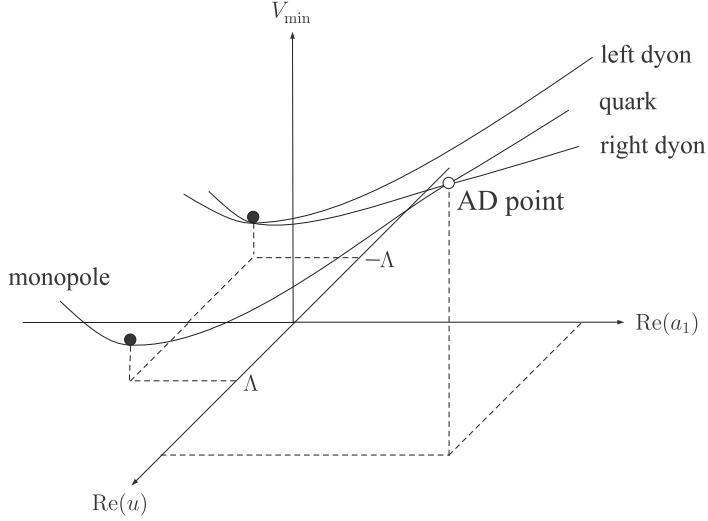}
\end{center}
\caption{Qualitative picture of the evolutions of the potential minima.}
\label{3d-plot}
\end{figure}

A qualitative picture of the evolutions of the potential minima
 is depicted in Fig. \ref{3d-plot}.

Now we have seen that there are two SUSY breaking minima and that
 there is no longer any runaway direction on the Coulomb branch.
It appears that the two SUSY breaking minima are global ones, but there
 is still a possible SUSY vacuum on the Higgs branch whose existence in
 the classical theory is shown in (\ref{Higgs-branch}).
It is known that there are no quantum corrections on the Higgs branch
 \cite{APS}.
Thus, at the quantum level, the SUSY vacuum on the Higgs branch
 is still left.
In the next section, we discuss the decay rate from the local SUSY breaking vacua at
 the monopole and dyon singular points to the SUSY vacuum on the Higgs
 branch and show that the local vacua can actually be meta-stable with an
 appropriate choice of parameters.

\section{Decay rate of the local vacuum \label{decay_rate}}

In this section, we estimate the decay rate from the SUSY breaking local minima
 on the Coulomb branch to the SUSY vacuum on the Higgs branch.

\begin{figure}[htb]
\begin{center}
\includegraphics[scale=.8]{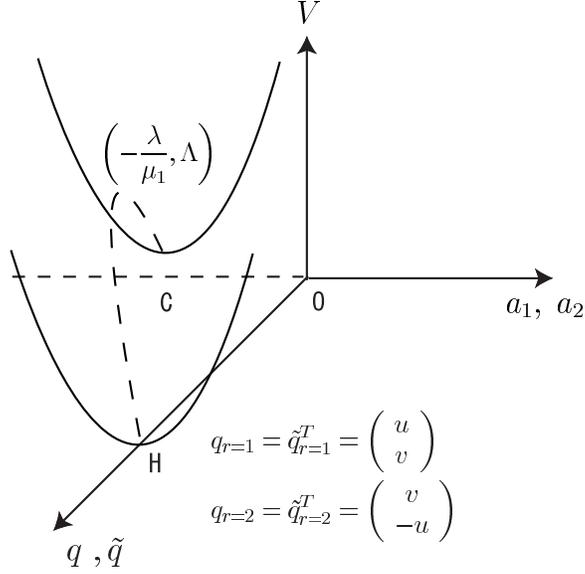}
\end{center}
\caption{Path from the Coulomb to Higgs vacuum}
\label{C-H}
\end{figure}

The local minimum at the monopole point on the Coulomb branch is approximately given
 by the point $C : (a_1, a_2) \sim (-\lambda/\mu_1, \Lambda), q = \tilde{q} = 0$ while
 the Higgs SUSY vacuum is
 at $H$, (\ref{Higgs-branch}) (see Fig. \ref{C-H}).
The distance between the vacua at Coulomb and Higgs branches,
 $|\overrightarrow{CH}|$,
 is given by
\begin{eqnarray}
|\overrightarrow{CH}|^2 = 4 (|u|^2 + |v|^2)
 + \left(\frac{\lambda}{\mu_1} \right)^2
 + \Lambda^2 \equiv L^2.
\end{eqnarray}
We parameterize a point between $C$ and $H$ by
 the vector
\begin{eqnarray}
\vec{p}(s) &\equiv&
\left(a_1, a_2, q_{r=1}, q_{r=2}, \tilde{q}_{r=1}, \tilde{q}_{r=2}
\right)
\nonumber \\
&=& s \left( \frac{\lambda}{\mu_1}, - \Lambda,
\left(
\begin{array}{c}
u \\
v
\end{array}
\right),
\left(
\begin{array}{c}
v \\
-u
\end{array}
\right),
\left(
\begin{array}{c}
u \\
v
\end{array}
\right),
\left(
\begin{array}{c}
v \\
-u
\end{array}
\right)
\right)
+ \left(
-\frac{\lambda}{\mu_1}, \Lambda, \mathbf{0}
\right)\,, \label{vec}
\end{eqnarray}
 where $0\le s\le 1$.
The parameter value $s = 0$ corresponds to the Coulomb vacuum
 while $s=1$ corresponds to the Higgs vacuum.
Substituting (\ref{vec}) into the classical potential
(\ref{tree_potential}), we have
\begin{eqnarray}
 V(s) = V(\vec{p}(s))
&\equiv& (1-s)^2(s^2\beta_1 + \beta_2)\,, \label{p1}
\end{eqnarray}
where
\begin{eqnarray}
& \beta_1 = 8 \left( \Lambda - \frac{\lambda}{\mu_1} \right)^2
 ({|u|^2 + |v|^2})
 + e^2 \lambda^2,~~~\beta_2 = 2 \mu_2^2 g^2 \Lambda^2\,.&
\end{eqnarray}

Now we show that there is a reasonable parameter region
 in which the local vacuum at the Coulomb branch, $C$, is meta-stable.
We take the following parameter region,
 $\Lambda\gg\lambda/\mu_1$ and $\lambda/\mu_2^2\gg g^2$,
 so that
\begin{eqnarray}
 {\beta_1 \over \beta_2}\sim {\lambda \over \mu_2^2 g^2}\gg 1\,,
  \label{barrier_condition}
\end{eqnarray}
where we have neglected the second term in $\beta_1$
 because of the small gauge coupling $e^2 \ll 1$.
Under eq. (\ref{barrier_condition}),
 the maximum value of the potential between $C$ and $H$ is located
 at $s=1/2$ and its value is given by
\begin{eqnarray}
\Delta V
= \frac{1}{16} \beta_1 + \frac{1}{4} \beta_2\,.
\end{eqnarray}
Since the SUSY breaking scale is estimated to be $V \sim \beta_2$,
we have
\begin{eqnarray}
\frac{\beta_2}{\Delta V} \sim \frac{\beta_2}{\beta_1+\beta_2}
\sim \frac{\beta_2}{\beta_1} \ll 1\,.
\end{eqnarray}
Thus we can use the thin-wall approximation to estimate the decay rate \cite{Co}.

The bounce action $B$ is evaluated in the triangle approximation \cite{DuJe}
\begin{eqnarray}
B = \frac{32 \pi^2}{3} \frac{(\Delta V_{+})^2 (\Delta \phi)^4}{\epsilon^3}
\end{eqnarray}
where $\epsilon \sim \beta_2, \ \Delta V_{+} = \Delta V - \epsilon, \
 \Delta \phi = L$ in our case.
The relevant
 quantities in the calculation are
\begin{eqnarray}
\frac{\Delta V_{+}}{\epsilon} \sim \frac{\beta_1}{\beta_2} \gg
 1\,,~~~~\frac{(\Delta \phi)^4}{\epsilon} \sim {\Lambda^2 \over g^2 \mu_2^2}\gg 1\,.
\end{eqnarray}
Then the bounce action is evaluated to be $B\gg 1$,
 and the decay rate $e^{-B}$ is extremely small.
Therefore, the SUSY breaking vacuum at the monopole point
 is meta-stable.
The decay rate from the degenerate dyon point to the SUSY vacuum
 and the one from the monopole point is the same due to the symmetric
 property in $u$-direction (see fig.\ref{3d-plot}).

\section{Conclusion and discussion \label{conclusion}}

We investigated an $SU(2) \times U(1)$ supersymmetric gauge theory
with $N_f=2$ massless flavors.
It contains soft terms, displayed in eq. (\ref{soft}), mass terms for $A_1$ 
and $A_2$ which break the $\mathcal{N} = 2$ SUSY down to $\mathcal{N} = 
1$ and a term (a Fayet-Iliopoulos term) linear in $A_1$.

We argued that when the parameters in the soft terms are small
 compared to the dynamical scale we can perform a reliable non-perturbative analysis
 based on the Seiberg-Witten solution.
Our analysis revealed an interesting setup:
On the Coulomb branch SUSY is dynamically broken in a manner reminiscent of the
 Izawa-Yanagida-Intriligator-Thomas model.
A local minimum emerges, but no runaway SUSY vacua survive.
On the Higgs branch, however, the SUSY vacua present at tree level
 should survive quantum corrections.
The local minimum on the Coulomb branch decays into the Higgs branch
 vacuum, but not surprisingly,
 the values of the parameters can be chosen such that it is very
 long-lived, {\it i.e.} meta-stable.

It is interesting to discuss the $U(1)_R$ symmetry.
In some class of models possessing a meta-stable SUSY breaking vacuum,
 an approximate $U(1)_R$ symmetry exists.
In our model, at the classical level the theory has
 an approximate $U(1)_R$ symmetry since we have taken the parameters in
 (\ref{soft}) to be small.
However, the $U(1)_R$ symmetry is broken to a discrete subgroup at the quantum level
even if there are no small superpotential perturbations.
Therefore, the theory does not have an approximate $U(1)_R$ symmetry
at the quantum level, so that the discussion in \cite{InSe} cannot be
applied to our case.

We would also like to comment on the difference between
 the models in this paper and in our previous paper \cite{ArMoOkSa}. 
Apart from the obvious difference that in this paper we start from a Lagrangian without extended SUSY, 
the pattern of vacuum states shows interesting differences:
In \cite{ArMoOkSa},
 the theory has SUSY vacua only on the Higgs branch while on the
 Coulomb branch the potential has pseudo flat directions at the
 classical level.
We found that after taking all the quantum corrections into account
 the effective potential exhibited a SUSY breaking local minimum.
In the present case, a SUSY vacuum exists on the Coulomb branch
 at the classical level.
We showed that the SUSY vacua are lifted by the gauge dynamics
 and revealed the mechanism how SUSY is dynamically broken.

In this paper, we chose a model simple enough to be able to perform a thorough analysis. As such, it is
too poor to serve as a basis for any realistic phenomenology. However, we think that
it once again shows the richness of supersymmetric gauge theories in being able to
provide instances of the most different kinds of properties, and in this respect we
hope that our simple model, like our previous attempt \cite{ArMoOkSa}, could provide
clues for building realistic descriptions of a world with broken supersymmetry.

\vspace{10mm}

\noindent {\large \bf Acknowledgements}

The work of N.~O. is partly supported by
the Grant-in-Aid for Scientific Research in Japan (\#15740164) and
 the Academy of Finland Finnish-Japanese Core Programme grant 112420. N.O. would also like to thank the High Energy Physics Division of the Department of Physical Sciences, University of Helsinki, for their hospitality during his visit.
S.~S. is supported by the bilateral program of Japan Society
for the Promotion of Science and Academy of Finland, ``Scientist
Exchanges.''\vspace{10mm}

\noindent {\Large \bf Appendix}
\begin{appendix}
\setcounter{equation}{0}
\def\theequation{A.\arabic{equation}}

\section{Explicit form of the effective couplings \label{appendix_A}}
In this appendix, we show the explicit
 forms of the periods $a(u), a_D(u)$,
 the effective couplings $b_{ij}$ and other quantities such as
 ${\partial u \over \partial a_i}$.
They are necessary for the
 analysis of the potential (\ref{potential2})
 since  the potential is a function of them.
A more detailed derivation of these expressions can be found in \cite{arai,ArMoOkSa}.
In the following, $\Lambda$ is a $SU(2)$ dynamical scale and
 $m$ is a common mass for the hypermultiplets, which is replaced with
 $a_1$ through the relation $m=\sqrt{2}a_1$ in the main body of the paper.

We first consider the periods $a_{2D}$ and  $a_2$.
Let us denote these as $a_{21}$ and $a_{22}$ respectively.
These are given by
\begin{eqnarray}
a_{2i}=-\frac{\sqrt{2}}{4\pi}
    \left(-\frac{4}{3}uI_1^{(i)}+8I_2^{(i)}
    +\frac{m^2\Lambda^2}{8}
    I_3^{(i)}
    \left(-\frac{u}{12}
    -\frac{\Lambda^2}{32}\right)\right)
    -\frac{m}{\sqrt{2}}\delta_{i2}\,,
    \label{eq:period2}
\end{eqnarray}
with the elliptic integrals
 $I_s^{(1)} \; (s=1,2,3)$ explicitly given by
\begin{eqnarray}
I_1^{(1)} &=&
           \frac{iK(k^\prime)}{\sqrt{e_2-e_1}}\,,
            \label{eq:formula1} \\
I_2^{(1)} &=&
            \frac{ie_1}{\sqrt{e_2-e_1}}K(k^\prime)
            +i\sqrt{e_2-e_1}E(k^\prime)\,,
            \label{eq:formula2} \\
I_3^{(1)} &=&
             = \frac{-i}{(e_2-e_1)^{3/2}}
            \left\{
            \frac{1}{k+\tilde{c}}K(k^\prime)
            +\frac{4k}{1+k}
            \frac{1}{\tilde{c}^2-k^{2}}
            \Pi_1\left(\nu,\frac{1-k}{1+k}
                \right)
            \right\}\,,
            \label{eq:formula3}
\end{eqnarray}
where $k^2 = \frac{e_3-e_1}{e_2-e_1}$,
      $k^{\prime 2}=1-k^2=\frac{e_2-e_3}{e_2-e_1}$,
      $\tilde{c}= \frac{c-e_1}{e_2-e_1}$,
and $\nu=-\left(\frac{k+\tilde{c}}{k-\tilde{c}}\right)^2
          \left(\frac{1-k}{1+k}\right)^2$.
Here
 $e_i(i=1,2,3)$ is a root of the elliptic curve for the ${\cal N}=2$ $SU(2)$
 QCD with massive $N_f=2$ flavors
\begin{eqnarray}
e_1 &=&\frac{u}{24}-\frac{\Lambda^2}{64}
     -\frac{1}{8}\sqrt{u+\frac{\Lambda^2}{8}
                  +\Lambda m}
                 \sqrt{u+\frac{\Lambda^2}{8}
                  -\Lambda m}\,,
                 \nonumber \\
e_2 &=&\frac{u}{24}-\frac{\Lambda^2}{64}
     +\frac{1}{8}\sqrt{u+\frac{\Lambda^2}{8} + \Lambda m}
                 \sqrt{u+\frac{\Lambda^2}{8} - \Lambda m}\,,
                 \label{eq:root}\\
e_3 &=&-\frac{u}{12}+\frac{\Lambda^2}{32}\,.
                 \nonumber
\end{eqnarray}
The formulae for $I_s^{(2)}$ are obtained from $I_s^{(1)}$
 by exchanging the roots $e_1$ and $e_2$.
In eqs.~(\ref{eq:formula1})-(\ref{eq:formula3}),
 $K$, $E$ and $\Pi_1$ are the complete elliptic integrals \cite{higher}
 given by
\begin{eqnarray}
K(k)&=&\int_0^1 \frac{dx}
     {\left[(1-x^2)(1-k^2x^2)\right]^{1/2}}\,,
       \\ \nonumber
E(k)&=&\int_0^1 dx\left(\frac{1-k^2x^2}{1-x^2}
     \right)^{1/2}\,,
       \\ \nonumber
\Pi_1(\nu,k)&=&\int_0^1\frac{dx}
     {[(1-x^2)(1-k^2x^2)]^{1/2}(1+\nu x^2)}\,.
\end{eqnarray}

Next let us consider the effective coupling defined in
 eq.~(\ref{effective_coupling}).
The effective couplings $\tau_{22}$ and $\tau_{12}$
 are obtained by
\begin{eqnarray}
\tau_{22}&=&\frac{\partial a_{2D}}{\partial
a_2}=\frac{\omega_1}{\omega_2}\,,\\
\tau_{12}&=&
\frac{\partial a_{2D}}{\partial a_1}
           \Bigg{|}_u
          -\tau_{22}
          \frac{\partial a_{2}}{\partial a_1}
           \Bigg{|}_u
=-\frac{2z_0}{\omega_2}\,,
         \label{eq:tau12}
\end{eqnarray}
where $\omega_i$ is the period of the Abelian differential,
\begin{eqnarray}
\omega_i
= 2I_1^{(i)} \; ~(i=1,2)\,,
\end{eqnarray}
and $z_0$ is defined as
\begin{eqnarray}
z_0=-\frac{1}{\sqrt{e_2-e_1}}F(\phi,k); \; \;
 \sin^2\phi=\frac{e_2-e_1}{c-e_1}\,.
\end{eqnarray}
Here $F(\phi,k)$ is the incomplete elliptic integral of
 the first kind given by
\begin{eqnarray}
F(\phi,k)=\int_0^{\sin\phi}\frac{dt}
          {[(1-t^2)(1-k^2t^2)]^{1/2}}\,.
\label{incomp}
\end{eqnarray}

The effective coupling $\tau_{11}$ is described
 in terms of the Weierstrass function
\begin{eqnarray}
\tau_{11}=-\frac{1}{\pi i}\left[\log\sigma(2z_0)
           +\frac{4z_0^2}{\omega_2}
           I_2^{(1)}\right] + C.
         \label{eq:tau11}
\end{eqnarray}
Here $\sigma$ is the Weierstrass sigma function,
 and $C$ is the constant in eq.~(\ref{eq:pre}).

\begin{figure}[htb]
\begin{center}
\includegraphics[scale=.8]{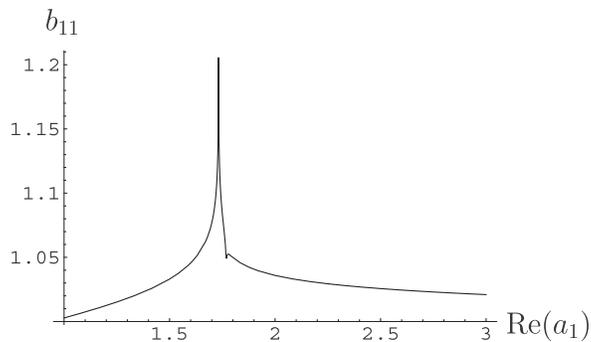}
\end{center}
\caption{Plot of the effective coupling $b_{11}$ as a function of ${\rm
 Re}(a_1)$ with $u=4$.}
\label{coupling}
\end{figure}

We now define the Landau pole associated with the $U(1)$ interaction.
In the ultraviolet region far away from the origin of the moduli space,
 the effective coupling is dominated by the $U(1)$ gauge interaction
 since the $SU(2)$ interaction is asymptotic free and small.
As expected, the gauge coupling $b_{11}$ is found to be
 a monotonically decreasing function of the large $|a_1|$
 with fixed $u$, and vice versa.
The Landau pole is defined as $|a_1|=\Lambda_L$ at which $b_{11}=0$.
The large $\Lambda_L$ required in our assumption
 is realized by taking an appropriate value for $C$.
In this paper, we fix $C =4 \pi i $,
 which corresponds to $\Lambda_L = 10^{17-18}$
 in units of $\Lambda$ \cite{arai,ArMoOkSa}.
Fig. \ref{coupling} shows the plot of the effective coupling $b_{11}$
 for $C=4\pi i$ as a function of ${\rm Re}(a_1)$ with $u=4$.
The cusps appear through the effect of the $SU(2)$ dynamics;
 their locations are specified by (\ref{descriminant}).

Finally we give the forms of ${\partial u \over \partial a_1}$
 and ${\partial u \over \partial a_1}$.
The former can be calculated as
\begin{eqnarray}
 {\partial u \over \partial a_1}{\Bigg |}_{a_2} &=& -{\partial u \over
  \partial a_1}{\Bigg |}_{a_2}{\partial a_2 \over \partial a_1}{\Bigg
  |}_{u}
 =-{1 \over \omega_2}{\partial a_2 \over \partial a_1}{\Bigg |}_u
  \nonumber \\
 &=& -{1 \over \pi}\left(\omega_2\zeta(z_0)-2z_0 \zeta\left(\omega_2
						       \over 2\right)\right)
 =-{m \Lambda^2 \over 16\pi}I_3^{(2)}\,.
\end{eqnarray}
The latter is simply given by
\begin{eqnarray}
 {\partial u \over \partial a_2}{\Bigg |}_{a_1}={1 \over \omega_2}\,.
\end{eqnarray}
\end{appendix}


\end{document}